\documentclass[12pt]{spieman}  
\usepackage{amsmath,amsfonts,amssymb}
\usepackage{graphicx}
\usepackage{setspace}
\usepackage{tocloft}
\usepackage[para,online,flushleft]{threeparttable}
\usepackage{soul}
\usepackage[dvipsnames]{xcolor}

\title{Optimization of EMCCD operating parameters for the acquisition system of SPARC4}

\author[a,b]{Denis Varise Bernardes}
\author[c,d]{Eder Martioli}
\author[a]{Danilo Henrique Spadoti}
\affil[a]{Universidade Federal de Itajubá, Laboratório de Telecomunicações - LabTel, Av. BPS, 1303 - Pinheirinho, Itajubá-MG, Brazil}
\affil[b]{Instituto Nacional de Pesquisas Espaciais, Avenida dos Astronautas, 1758, São José dos Campos-SP, Brazil}
\affil[c]{Sorbonne Université, CNRS, UMR 7095, Institut d’Astrophysique de Paris, 98 bis bd Arago, 75014 Paris, France}
\affil[d]{Laboratório Nacional de Astrofísica, Rua Estados Unidos, 154, Itajubá-MG, Brazil}

\cftpagenumbersoff{figure}
\cftpagenumbersoff{table} 
\begin{document} 
\maketitle


\begin{abstract}
We present the Optimization Method for the Electron Multiplying Charge Coupled Devices (EMCCDs) of the Acquisition System of the SPARC4 (OMASS4). The OMASS4 uses as figures of merit the signal-to-noise ratio (SNR) and the acquisition rate (AR) as a function of the operation mode of the CCDs. Three different modes of optimization are included in the OMASS4: (1) optimization of SNR only; (2) optimization of AR only; and (3) optimization of both SNR and AR simultaneously. The first two modes calculate an analytical maximization of the cost function whereas the third mode uses the bayesian optimization method to determine the optimum mode of operation. We apply the OMASS4 to find the optimum mode for observations obtained at the Pico dos Dias Observatory, Brazil, and compare the delivered modes of operation and its performance with the ones adopted by the observer. If the OMASS4 had been used as a tool to optimize the CCDs in all of these nights, it would be possible to improve their efficiency in 97.17 \%, 65.08 \%, and 77.66 \% for the optimization modes 1, 2, and 3, respectively. 

\end{abstract}


{\noindent \footnotesize\textbf{*}Denis Varise Bernardes: \linkable{denis.bernardes@inpe.br} \newline}
{\noindent \footnotesize\textbf{*}Eder Martioli:  \linkable{emartioli@lna.br} \newline}
{\noindent \footnotesize\textbf{*}Danilo Henrique Spadoti:  \linkable{spadoti@unifei.edu.br}}


\section{Introduction} \label{sec:introduction}

The Astrophysics Division of the {\it Instituto Nacional de Pesquisas Espaciais} (INPE) in collaboration with the {\it Laboratório Nacional de Astrofísica} (LNA) is developing a new astronomical instrument, the Simultaneous Polarimeter and Rapid Camera in Four Bands (SPARC4) \cite{Rodrigues2012}. SPARC4 will be installed on the 1.6~m Perkin-Elmer telescope at Observatório Pico dos Dias, Brazil, and it will allow image acquisition in the four Sloan Digital Sky Survey (SDSS) photometric bands: g, r, i and z~\cite{SDSS}. The synchronization of the acquisition of these photometric bands (channels) will be done by using a digital pulse generator developed by the Highland Technology Company \cite{PulseGenerator}. Different synchronization modes for the four channels can be implemented by programming the pulses created by the pulse generator. Also, the SPARC4 will allow image acquisition using the photometric or polarimetric mode. In the polarimetric mode, two wave plates are introduced in the light path that allow measurements with linear and circular polarimetry.

For the acquisition in each channel, there is a dedicated iXon Ultra EMCCD, produced by Andor Technology \cite{ixon_hardware_guide}. These devices have an optical window and coating optimized for the spectral range in which they were designed to operate \cite{Rodrigues2012}. These cameras also have frame transfer and electron-multiplying capabilities, allowing acquisition rates (AR) of up to 26~fps full-frame (1024 x 1024 pixels) even on faint astronomical objects, which requires high sensitivity for short exposure times.   

The quality of photometric measurements in astronomical observations can be quantified by the signal-to-noise ratio (SNR). Another important constraint concerning many scientific applications expected with SPARC4, especially those requiring fast time-series photometry or polarimetry, is the AR.  Either the SNR or AR or both can change depending on the configuration of the operational mode of the CCD. Therefore, an optimal selection of the operational mode for each CCD is important to obtain the best performance of the instrument. The CCDs provide a set of parameters to control the operational modes, such as the horizontal and vertical shift speed, the CCD gain, and the electron multiplying on/off mode. In addition, one may consider other parameters to obtain an optimal performance such as the spatial binning, exposure time, and sub-imaging, all of which may or may not have their values restricted by the scientific requirements. These parameters affect the SNR and AR in different ways, with non-linear dependencies, and with additional restrictions. Therefore, an optimal parameters choice for all four CCDs to obtain the highest performance in observations may be a difficult task for a human, even for experienced and skillful observers. A solution to this problem concerns the main objective of this work. 

In this scenario, Ref.~\citenum{Wu2010a} derived the SNR for EMCCDs to estimate the resolution of a microscope as a function of the EM gain $G_{\rm em}$. The authors state that there is a critical value of $G_{\rm em}$ for which the SNR starts to degrade. The authors in Ref.~\citenum{Shim2014} presented a method to auto-adjust the exposure time $t_{\rm exp}$ to optimize the information acquired by the cameras in different lighting conditions. This method was based on the calculation of a gradient for a series of images taken at different brightness intensities. The optimum $t_{\rm exp}$ is the one that maximizes the gradient. The methodology presented in Ref.~\citenum{Wang2012}, and Ref.~\citenum{Lu2010} was based on the maximization of the Shannon entropy. To accomplish this, Ref.~\citenum{Wang2012} adjust the $t_{\rm exp}$ of the cameras. Ref.~\citenum{Lu2010}, in turn, acquire a sequence of images with different combinations of $t_{\rm exp}$ and gain. With these experiments the authors stated that there was an optimum combination for both parameters and the optimum values varying according to the illumination conditions. In Ref.~\citenum{Kim2018a} the authors presented a metric that assesses the image quality according to the $t_{\rm exp}$ and gain values. The optimum configurations for both parameters were found through the application of the Bayes optimization method to this metric. In this way, Ref.~\citenum{Bergstra2011} presented a methodology to optimize a 32-dimensional (including discrete and continuous variables) Deep Belief Networks using the Bayes optimization method with the Tree-structured Parzen Estimator (TPE) algorithm. The authors obtained the best result in hyper-parameter optimization of Deep Belief Networks, outperforming previous results obtained by using human-guided search, brute-force random search, and the Gaussian Process algorithm. Among the authors, only Ref.~\citenum{Bergstra2011} have presented a methodology to optimize several parameters simultaneously, which are discrete and continuous. In our case, there is also a mix of discrete and continuous parameters for the configuration of the SPARC4 cameras. 

Inspired on the above methodologies we consider using the Bayes Optimization Method, which is designed for black-box derivative-free global optimization~\cite{Frazier2018} to find the optimal values of relevant configuration parameters of the CCDs that provides the highest SNR and/or AR. This is the first step to optimize the performance of SPARC4. This paper is organized as follows. Section \ref{sec:BOM} presents the Bayes optimization method used in this work. Section \ref{sec:SNR} presents the considerations adopted for the SNR calculation. Section \ref{sec:acquisition_system} presents the SPARC4 acquisition system and the characterization of one of the detectors. Section \ref{sec:opt_algoritmo} describes the OMASS4. Section \ref{sec:results} presents the results and discussion, and Sec. \ref{sec:conclusion} presents the conclusion.


\section{Bayes optimization method} \label{sec:BOM}

The Bayes optimization method is a statistical free derivative method to search the maximum or minimum of a non-linear objective function $f(x)$ using a small data set \cite{Frazier2018},  \cite{Berk2019}. The execution of the Bayes optimization method can be divided into two parts: the estimation of the objective function through a bayesian statistical model and the application of an acquisition function to determine the next best point to evaluate \cite{Brochu2010a}. 

In estimating the objective function one obtains a noisy sample for each iteration of the Bayes optimization method. These samples provide a prior distribution for the objective function, which are combined with a likelihood function for the samples obtained until that moment, resulting in a posterior distribution. This step can be interpreted as estimating the objective function through a surrogate function \cite{Brochu2010a} \cite{Brochu2010}. Some of the most known surrogate functions are the Gaussian Process (GP)~\cite{Frazier2018}, the Random Forest Regression \cite{Dewancker2015}, and the TPE~\cite{Bergstra2011}. Between them, the TPE was chosen to be used in OMASS4, allowing the optimization of continuous and discrete parameters simultaneously. Also, the objective function used by the OMASS4 is presented in the Eq. \ref{eq:funcao_custo_SNR_FA} and it will be discussed later.

In execution of the acquisition function part, the goal is to determine the next value to evaluate that is the most likely optimum point of the objective function. Thus, one must look for points with high probability to be the next optimal point, points with high uncertainty, or both at the same time \cite{Brochu2010a} \cite{Brochu2010}. This step is known as the exploitation/exploration balance. The acquisition function used in Ref. \citenum{Bergstra2011} to implement the TPE is the Expected Improvement, as described in the next section.

\subsection{Tree-Structured Parzen Estimator Algorithm (TPE)}

The TPE algorithm was firstly presented by Ref. \citenum{Bergstra2011}, and it was used to obtain the minimum of a function through an interactive fit of a probability density function $p(y|x)$, where $y$ is the value of the evaluation of objective function in $x$ \cite{Kastner.2019}. The TPE uses the Bayes rule in the form

\begin{equation}
    p(y|x) = \frac{p(x|y) \times p(y)}{p(x)},
\label{eq:regra_bayes}    
\end{equation}

\noindent where $p(x|y)$ is the probability of $x$ to be chosen as the next evaluation point \cite{Singh2018}. In turn, $p(x|y)$ can be expressed as 

\begin{equation}
    p(x|y) = 
    \Bigg \{
    \begin{array}{cc}
         l(x), & if \; y < y^* \\
         g(x), & if \; y \geq y^*.
    \end{array}
\label{eq:lx_gx_TPE}    
\end{equation}

\noindent $l(x)$ is the density probability function given by those points which $f(x) < y^*$ and $g(x)$ is the probability density function given by the other points. The value of $y^*$ is fit by the algorithm over the iterations, so that a percentage $\gamma$ of the data accomplish the condition $p(y<y^*) = \gamma$ \cite{Dewancker2015} \cite{Singh2018}. Thus, the expression for the expected improvement $E_{\rm I}$ for the TPE acquisition function is given as follows

\begin{equation}
     E_{\rm I} = \frac{\gamma y^* l(x) - l(x) \int_{\infty}^{y^*} p(y)dy}{ \gamma l(x) + (1-\gamma)g(x)}\\
      \propto \Bigg [\ \gamma + \frac{g(x)}{l(x)}(1-\gamma)  \Bigg ]\ ^{-1}. \\
\label{eq:ME_TPE_4}    
\end{equation}

\noindent The deduction of the Equation \ref{eq:ME_TPE_4} can be found in Ref. \citenum{Bergstra2011}. This function demonstrates that the maximization of the $E_{\rm I}$ occurs for those points $x$ in the domain of objective function with high probabilities for $l(x)$ and low probabilities for $g(x)$.


\section{Signal-to-noise ratio} \label{sec:SNR}

\begin{figure}[h!]
    \centering
    \includegraphics[scale=0.75]{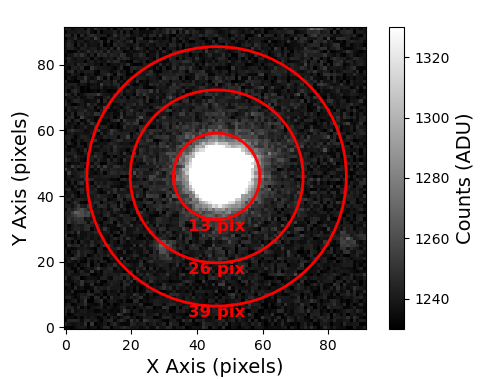}
    \caption{Counts distribution over the pixels of the CCD. The image presents three concentric circles: with 13 pixels radius, 26 pixels radius, and 39 pixels radius. Pixels in the inner circle are those considered for calculation of the star flux. Pixels between circles with radius 26 pixels and 39 pixels are those considered for the calculation of the background flux.}
    \label{fig:estrela_circulos}
\end{figure}

In this scope it will be considered as a signal the photon flux measured by the CCD integrated over time and space. Figure \ref{fig:estrela_circulos} illustrates an example of the flux of a point-like source (e.g., a star) acquired by a CCD distributed over the pixels, where the three concentric circles have radius of 13, 26, and 39 pixels. Only pixels lying within the inner circle are considered for the calculation of the star flux $S$. Pixels lying within the outer annulus are considered for the calculation of the background flux $S_{\rm f}$. The radius of these circles was obtained by modeling the star flux distribution as a Gaussian distribution, so the star radius was calculated through the full width at half maximum (FWHM) \cite{FWHM} parameter. The star radius equals $3 \times$ the circle radius at FWHM. The second and the third circles have two and three times the radius of the inner circle, respectively. 

Eq. \ref{eq:snr_1} presents the SNR $\mathcal{S}$ calculation of the Fig. \ref{fig:estrela_circulos}

\begin{equation}
    \mathcal{S} = \frac{S}{N},
\label{eq:snr_1}    
\end{equation}

\noindent where $N$, in electrons, is the total noise of the image. $S$, in electrons, is given by the sum of the pixels values $S_i$ in the inner circle, in the analogical-to-digital unit (ADU), subtracted by $S_{\rm f}$~\cite{Martioli2018}, as follows:

\begin{equation}
    S = \sum_{i=1}^{n_{\rm p}} ( S_i - S_{\rm f} ) \times G.
\label{eq:sinal_estrela}    
\end{equation}

\noindent $n_{\rm p}$ is the number of pixels considered to calculate the star flux, $S_{\rm f}$ is given by the median, in ADU, of the pixels in the annulus and $G$ is the conversion factor, or gain, of the CCD, in e-/ADU. 

According to Ref. \citenum{Merline1995}, it can be said that $N$ is composed by the read noise (RN) $\sigma_{\rm r}$, the dark current (DC) noise $\sigma_{\rm dc}$, the noise of sky measurement $\sigma_{\rm sky}$ and the noise of star flux measurement $\sigma_{\rm s}$. There are other noise sources in the CCD, but, for simplification, they will not be considered in this scope. $\sigma_{\rm r}$ is the standard deviation, given in e- rms, around the bias level of the CCD. $\sigma_{\rm dc}$ is given in electrons and can be calculated through the mean of the electrons per pixel $S_{\rm dc}$ generated by the DC as a function of the $t_{\rm exp}$, where $S_{\rm dc}$ is given as

\begin{equation}
    S_{\rm dc} = D_{\rm C} \times t_{\rm exp},
\label{eq:corrente_escuro}    
\end{equation}

\noindent where $D_{\rm C}$ is the dark current, given in e-/pixel/s, for the respective CCD temperature \cite{DenisVB}. $\sigma_{\rm sky}$ is given in photons and can be obtained through the mean of the photon flux per pixel of the sky $S_{\rm sky}$:

\begin{equation}
    S_{\rm sky} = (S_{\rm f} - B) \times G - S_{\rm dc}, 
\label{eq:calc_fluxo_ceu}    
\end{equation}

\noindent where $B$ is the bias level of the image, in ADU. $\sigma_{\rm s}$ is given in number of photons, and it can be calculated through the star flux $S$. $\sigma_{\rm dc}$, $\sigma_{\rm sky}$, $\sigma_{\rm s}$ can be calculated assuming a Poisson distribution \cite{Merline1995}. Thus, the expression for each noise source is

\begin{equation}
    \sigma_{\rm r} = \sum_{i=1}^{n_{\rm p}} (\sigma_{\rm ADU (i)} \times G)
\label{eq:Poisson_1}    
\end{equation}

\begin{equation}
    \sigma_{\rm dc}^2 = \sum_{i=1}^{n_{\rm p}} S_{\rm dc (i)} \times G_{\rm em}^2  \times N_{\rm F}^2
\label{eq:Poisson_2}    
\end{equation}

\begin{equation}
    \sigma_{\rm sky}^2 = \sum_{i=1}^{n_{\rm p}} S_{\rm sky(i)} \times G_{\rm em}^2  \times N_{\rm F}^2
\label{eq:Poisson_3}    
\end{equation}

\begin{equation}
    \sigma_{\rm s}^2 = S \times G_{\rm em}^2  \times N_{\rm F}^2,
\label{eq:Poisson_4}    
\end{equation}

\noindent where $\sigma_{\rm ADU (i)}$, $S_{\rm dc (i)}$, $S_{\rm sky(i)}$ represent the counts' distribution per pixel, the mean thermoelectrons per pixel, and the photons number per pixel of the sky, for the i-th pixel of the star, respectively. $G_{\rm em}$ is the amplification gain provided by the EM mode of the CCD. For the conventional mode, $G_{\rm em}$ = 1. $N_{\rm F}$ is the noise factor and represents and extra noise added to the image because of the use of the EM amplifier. For an Andor EMCCD, $N_{\rm F}$ = 1.41 \cite{Andor_max_emgain}. However, it will be considered that these parameters are constant over the pixels, so

\begin{equation}
    \sigma_{\rm r} = n_{\rm p} \times \sigma_{\rm ADU} \times G
\label{eq:Poisson_5}    
\end{equation}

\begin{equation}
    \sigma_{\rm dc}^2 = n_{\rm p} \times S_{\rm dc} \times G_{\rm em}^2  \times N_{\rm F}^2
\label{eq:Poisson_6}    
\end{equation}

\begin{equation}
    \sigma_{\rm sky}^2 = n_{\rm p} \times S_{\rm sky} \times G_{\rm em}^2  \times N_{\rm F}^2. 
\label{eq:Poisson_7}    
\end{equation}

By considering uncorrelated noise sources \cite{Newberry_1991}, it is possible to express $N$ as follows

\begin{equation}
    N^2 = \sigma_{\rm s}^2 +  \sigma_{\rm sky}^2 + \sigma_{\rm dc}^2 + \sigma_{\rm r}^2.
\label{eq:ruido_total_1}    
\end{equation}

Replacing the Eq. \ref{eq:Poisson_4} - \ref{eq:Poisson_7} on Eq. \ref{eq:ruido_total_1}, we find

\begin{equation}
        N^2 = \\
         S \times G_{\rm em}^2  \times N_{\rm F}^2  + \\
         n_{\rm p} \times [\ S_{\rm sky} +  S_{\rm dc} \times G_{\rm em}^2  \times N_{\rm F}^2] + \\
         n_{\rm p} \times (\sigma_{\rm ADU} \times G)^2.
\label{eq:ruido_total_2}    
\end{equation}

Replacing equation Eq. \ref{eq:ruido_total_2} on Eq. \ref{eq:snr_1}, and dividing the fraction by $G_{\rm em}$, the SNR will be

\begin{equation}
        \mathcal{S} = \\
        \frac{S}{\sqrt{S \; N_{\rm F}^2  + n_{\rm p} \; [\ (S_{\rm sky} +  S_{\rm dc}) \; N_{\rm F}^2   +  (\sigma_{\rm ADU} \; G/G_{\rm em})^2 ]\ }}.
\label{eq:SNR_2}    
\end{equation}

Thus, we find an expression for the SNR that is adopted in the OMASS4. Given the information of the star flux and the image noise, this equation was used to estimate the SNR values for different operation modes of the CCDs and then to determine the optimum mode.


\section{Acquisition System} \label{sec:acquisition_system}

For the control of the SPARC4 acquisition system, a software prototype was developed in Labview language~\cite{LabVIEW} using the Software Development Kit \cite{SDK} developed by Andor Technology for communication with the CCDs. This software will allow simultaneous and synchronized acquisition for the four SPARC4 channels. The synchronization will be made by a digital pulse generator \cite{PulseGenerator} developed by the Highlands Technology with a resolution of 10~ps between pulses. For each channel it is possible to acquire cubes with up to 70~full-frame images with a delay of approximately 1.7~ms between exposures. It is also possible to concatenate cubes with a delay between 160~ms to 980~ms depending on the size of the cube. Thus, this system will allow the acquisition of synchronized image cubes for the four channels, a feature that is not available on the control software delivered by the manufacturer, the Andor Solis~\cite{AndorSolis}.

The critical parameters that are employed in our system to control the cameras are: CCD temperature, $G_{\rm em}$ mode, sub-image (SI), and binning of the pixels (Bin).  Their allowed values are constrained either by the scientific application or to ensure the safety of the system. The CCD temperature, for instance, is constrained in the interval between -80~ºC and 20~ºC~\cite{ixon_hardware_guide}. The $G_{\rm em}$ is limited to the range between 2$\times$ and 300$\times$, which is the safe operational range advised by the manufacturer \cite{Andor_max_emgain}. The SI is limited to the same set of options available in the Andor Solis, i.e., squared windows with sizes of 256~pixels, 512~pixels, and 1024~pixels. Finally, the electronic binning was limited to the values of 1~$\times$~1 pixel and 2~$\times$~2 pixels. Given that the SPARC4 plate scale is 0.35~arcsec/pixel~\cite{SPARC4_2012} and the median seeing at Observatório Picos dos Dias is 1.5~arcsec~\cite{Carvalho2011}, the choice of using only these two options for binning values is to ensure that the spatial resolution of the instrument will not be affected. Table \ref{tab:parametros_controle_CCD} presents all the control parameters of the acquisition system and their respective ranges. Based on these criteria, we characterized the read noise and the acquisition rate of one of the SPARC4 CCDs for all of the allowed operation modes.

\begin{table}[h!]
\centering
\caption{Control parameters of the operation mode of the CCDs made available by the acquisition system. For each parameter, the unit and allowed values are provided}.
\label{tab:parametros_controle_CCD}
\begin{tabular}{ccc}
\hline
Parameter & Unit &  Allowed values \\
\hline
Exposure time & (s) &  $>$ 1$\times 10^{-5}$\\
Images in cube & &  $>$ 1, and $<$ 70\\
CCD cooler & & on/off\\
CCD temperature & (ºC) & from -80 to 20\\
Horizontal Shift Speed & (MHz) & 0,1; 1; 10; 20 and 30\\
Pre-amplification & & 1 and 2\\
EM mode & & on/off\\
EM gain & & from 2x to 300x\\
Sub-image & (pixels) & 256,  512, and 1024\\
Binning & (pixels) & 1 and 2\\
\hline
\end{tabular}
\end{table}

\subsection{Characterization of the CCDs} \label{subsec:charact_CCDs}

This section presents the experimental procedures and results obtained in the determination of the RN and the AR for all available modes of operation for one of the SPARC4's detectors, the I channel, an EMCCD iXon Ultra 888, serial number 9916. Based on these results, two packages were developed using Python language. One package was developed using the characterization of the read noise presented in Sec. \ref{subsec:read_noise_char} to calculate the SNR as a function of the operation mode of the CCD. The other one was developed using the results presented in Table \ref{tab:tempo_critico_all_modes} of the Sec. \ref{subsec:frequency_acq_char} for the calculation of the AR as a function of the operation mode. These packages, in turn, are used by the OMASS4 to determine the optimum operation mode of the CCDs. 

\subsubsection{Read Noise} \label{subsec:read_noise_char}

This section presents the characterization of the RN, implemented based on the methodology presented in Ref. \citenum{DenisVB} for the SPARC4 cameras. The characterization was performed for all combinations of the allowed values for the parameters Horizontal Shift Speed (HSS), pre-amplification (PA), and Bin. For the EM mode, the following $G_{\rm em}$ values were used: 2x; the range 10x to 50x, in steps of 10x; and the range 50x to 300x, in steps of 50x. For each evaluation, we calculated the standard deviation of the counts for each pixel, for a series of 100 bias images. This procedure resulted in an image of the spatial distribution of the EMCCD noise for the respective operation mode. From this image noise, we calculated the probability distribution function of the pixels. Therefore, the RN presented in this characterization is the median of this probability distribution, multiplied by the CCD gain, for the conversion from counts to electrons. The RN error is calculated by the absolute standard deviation of the probability distribution, also, multiplied by the CCD gain.

We estimated the DC for the used EMCCD using the model presented in Ref. \citenum{DenisVB} for the SPARC4 cameras:

\begin{equation}
    DC_{\rm 9916} = 9.67 \; e^{0.0012 T^2 + 0.25 T},
\end{equation}

\noindent where $DC_{\rm 9916}$ is the model that estimates the DC of the 9916 camera for the temperature range -30~ºC to -70~ºC. T = -60 ºC is the used temperature for the EMCCD. Therefore, the calculated DC is 3.93~$\times$~10$^{-8}$~e-/pix/s. In the worst scenario, for the minimum exposure time of 1~$\times$~10$^{-5}$~s and the maximum $G_{\rm em}$ value of 300x, the DC noise is $\sim$1.18~$\times$~10$^{-10}$~e-/pix. For this reason, the DC noise was neglected in this analysis.

\begin{figure*}[h!]
    \centering
    \includegraphics[scale=0.75]{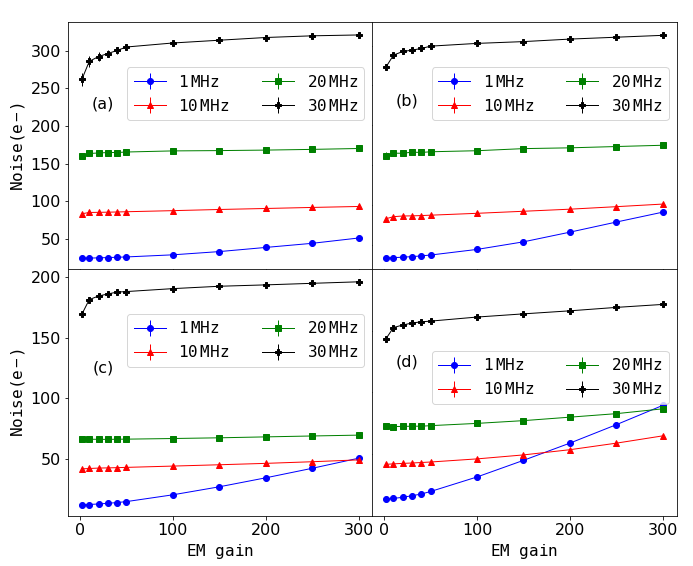}
    \caption{Read noise of the CCD as a function of the EM gain. In panels, it is presented the results obtained for the Horizontal Shift Speed modes 1 MHz (blue), 10 MHz (red), 20 MHz (green), and 30 MHz (black). The used values for pre-amplification (PA) and binning (Bin) for each panel are (a) PA = 1, Bin = 1 pixel; (b) PA = 1, Bin = 2 pixel; (c) PA = 2, Bin = 1 pixel; (d) PA = 2, Bin = 2 pixels}.
    \label{fig:EM_x_RN}
\end{figure*}

The RN values obtained for the conventional mode are presented in Table \ref{tab:rn_conv}. Values without errors were obtained from the camera data sheet~\cite{iXonDataSheet}. Figure \ref{fig:EM_x_RN} presents the RN values obtained for the EM mode characterization. Each point in the plot represents the mean and standard variation of three measurements. From a given $G_{\rm em}$ value, the noise for the 1 MHz mode is greater than the noise for the modes 10 MHz (Figure \ref{fig:EM_x_RN}c and \ref{fig:EM_x_RN}d) and 20 MHz (Figure \ref{fig:EM_x_RN}d). This is unexpected behavior. We believe that it would be a noise created by the electronic of this specific device, but further investigations involving the other SPARC4 cameras are needed to confirm this statement. For now, this noise was taken into account, and it decreases, in some cases, the real value of the SNR. However, the OMASS4 operation does not depend on the SNR profile, and it still manages to find the optimum operation mode, as presented in Section \ref{subsubsec:opt_mode_validation}.

\begin{table}[h!]
\centering
\caption{Read noise values obtained for the characterization of the conventional mode of the EMCCD. Column Error presents the standard deviation of the read noise measurements in electrons. The read noise values without error were obtained from the camera datasheet~\cite{iXonDataSheet}}.
\label{tab:rn_conv}
\begin{tabular}{ccccc}
\hline
HSS & PA & Binning & RN & Error \\
(MHz) & & & (e-) & (e-) \\
\hline
1 & 1 & 1 & 6.67 & \\	
1 & 1 & 2 & 6.94 & 0.22\\
1 & 2 & 1 & 4.76 &\\
1 & 2 & 2 & 4.79 & 0.93\\
0.1 & 1 & 1 & 8.78 & \\	
0.1 & 1 & 2 & 8.84 & 0.22\\
0.1 & 2 & 1 & 3.46 &\\
0.1 & 2 & 2 & 3.27 & 0.93\\
\hline
\end{tabular}
\end{table}

\subsubsection{Acquisition Rate characterization} \label{subsec:frequency_acq_char}

\begin{figure}[h!]
    \centering
    \includegraphics[scale=0.6]{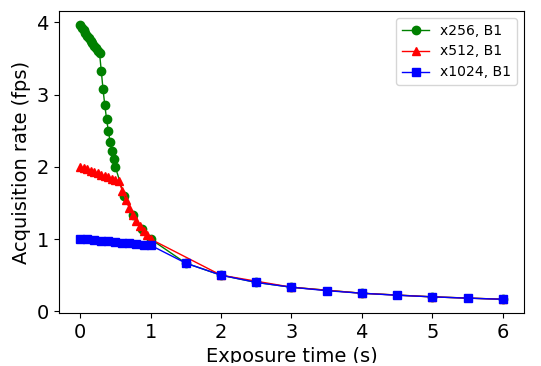}
    \caption{Acquisition rate as a function of the exposure time for the mode Horizontal Shift Speed of 1 MHz and Binning of 1~pixel. The colour code of curves are: Sub-images 256x256 pixels (green), 512x512 pixels (red), and 1024x1024 pixels (blue)}.
    \label{fig:AR _vs_texp}
\end{figure}

A characterization of the AR for every mode allowed by the SPARC4 for parameters $t_{\rm exp}$, HSS, Bin, and SI was performed. For each mode, the AR was calculated as a function of the $t_{\rm exp}$. For each $t_{\rm exp}$ value, measurements were made by acquiring a cube with 10 images. Thus, the AR  was calculated by dividing the number of images by the time taken to acquire the cube. Measurements were repeated three times and the AR represents the mean value. Figure \ref{fig:AR _vs_texp} presents, as an example, the result obtained for the mode HSS~=~1~MHz, Bin~=~1~pixel, for all SI values. According to Fig. \ref{fig:AR _vs_texp}, it is possible to notice two regimes for the AR: a linear regime and a decaying regime. This occurs because of the frame transfer option of the CCD, in which an image is read in parallel with the acquisition of the next image~\cite{ixon_hardware_guide}. Linear regime occurs for $t_{\rm exp}$ values smaller than the critical time $t_{\rm c}$ for that mode. The decaying regime occurs for $t_{\rm exp}$ values larger than the $t_{\rm c}$, where the AR is given by the inverse of $t_{\rm exp}$. $t_{\rm c}$ is composed of the image readout time plus internal delays. The inflection points presented in Fig. \ref{fig:AR _vs_texp} provide an estimate of the $t_{\rm c}$. Thus, a linear and a $1/t_{\rm exp}$ function fit were performed for each of the two regimes, and $t_{\rm c}$ was obtained at the intersection of these two. This procedure was repeated for every mode of the SPARC4, and the result is presented in Table \ref{tab:tempo_critico_all_modes}.

\begin{table}[h!]
\centering
\caption{Critical time values $t_{\rm c}$ of the CCD, for each operation mode given by the Horizontal Shift Speed (HSS), Sub-image (SI) and Binning parameters.}
\label{tab:tempo_critico_all_modes}
\begin{tabular}{@{}cccc}
\hline
HSS & SI & Binning & $t_{\rm c}$ \\
(MHz) & (pixels) & (pixels) & (s) \\   
\hline
    30  & x256  & 2 & 0.0059   \\
    20  & x256  & 2 & 0.0083  \\
    30  & x256  & 1 & 0.011   \\
    30  & x512  & 2 & 0.011  \\
    20  & x256  & 1 & 0.015  \\
    20  & x512  & 2 & 0.015  \\
    10  & x256  & 2 & 0.015  \\
    30  & x512  & 1 & 0.020  \\
    30  & x1024 & 2 & 0.020  \\
    10  & x256  & 1 & 0.029  \\
    20  & x512  & 1 & 0.029  \\
    10  & x512  & 2 & 0.029  \\
    20  & x1024 & 2 & 0.029  \\
    30  & x1024 & 1 & 0.039  \\
    10  & x512  & 1 & 0.057  \\
    10  & x1024 & 2 & 0.057  \\
    20  & x1024 & 1 & 0.057  \\
    10  & x1024 & 1 & 0.11   \\
    1   & x256  & 2 & 0.14  \\
    1   & x256  & 1 & 0.28  \\
    1   & x512  & 2 & 0.28  \\
    1   & x1024 & 2 & 0.56  \\
    1   & x512  & 1 & 0.56  \\
    1   & x1024 & 1 & 1.11  \\
    0.1 & x256  & 2 & 1.24  \\
    0.1 & x512  & 2 & 2.14  \\
    0.1 & x256  & 1 & 2.79  \\
    0.1 & x1024 & 2 & 2.97   \\
    0.1 & x512  & 1 & 5.53  \\
    0.1 & x1024 & 1 & 10.93 \\
\hline
\end{tabular}
\end{table}


\section{Optimization Method for the EMCCDs of the Acquisition System of SPARC4} \label{sec:opt_algoritmo}

An algorithm was developed using the Python Language 3.7.4 to determine the optimum operation mode of the SPARC4 CCDs. This algorithm was structured into three parts: the initialization, the star flux calculation, and the CCD optimization. For the initialization step, it requires to provide to the software all the information related to the astronomical object, i.e.: an image of the object, its $x,y$ coordinates, the maximum star radius, a bias image with the respective used CCD operation mode, the SNR, the AR, allowed SI and Bin modes, CCD temperature, and the iterations number of the Bayes optimization method. Then, the star flux is calculated for the optimal star radius given by the FWHM parameter, as described in Sec.~\ref{sec:SNR}. The OMASS4 uses a set of packages to calculate the SNR and the AR values according to the CCD operation mode. The code developed to execute the Bayes optimization method is based on the library provided by Ref. \citenum{hyperopt_koehrsen}. The TPE algorithm was used to model the objective function of the Bayes optimization method. The SNR package operation is based on the methodology presented in Sec.~\ref{sec:SNR}. The star flux, sky flux, and the number of star pixels are obtained in the previous step. The DC noise is calculated according to the model presented by Ref.~\citenum{DenisVB} for the four SPARC4 cameras. The read noise is obtained through the characterization presented in Sec.~\ref{subsec:read_noise_char}. For conventional modes, the values in Table \ref{tab:rn_conv} were used. For EM modes, an interpolation of the data presented in Fig. \ref{fig:EM_x_RN} was made. The $G$ value was obtained from the camera datasheet \cite{iXonDataSheet}.

The performance of the EM mode is better than the conventional mode until a maximum value of 100 photons per pixel \cite{Andor_max_emgain}. So, the $t_{\rm exp}$ of each EM mode is limited to accomplish this requirement. Also, the maximum value allowed for the $G_{\rm em}$ is 300x. Values larger than 300x would deteriorate the device~\cite{AndorSolis}. Furthermore, the $G_{\rm em}$ must be such that the CCD will not saturate. For this reason, the maximum EM gain allowed was arbitrarily configured to provide a signal up to 80 \% of the pixel well depth. For an image with 16 bits per pixel, this value is $2^{16} \times 0.8 \approx 52429$ ADU. Given that a pixel value is composed by the star, sky and, dark current signals, and the bias level, the maximum value for the $G_{\rm em}$ is

\begin{equation}
    G_{\rm em} = \frac{(52429 - B) \times G}{(S/n_{\rm p} + S_{\rm sky} + S_{\rm dc})}.
\label{eq:max_em_gain}    
\end{equation}

The AR package operation is based on the characterization presented in Sec. \ref{subsec:frequency_acq_char}. For each mode, the AR value is calculated using linear interpolation if $t_{\rm exp} < t_{\rm c}$, and $1/t_{\rm exp}$, if $t_{\rm exp} \geq t_{\rm c}$. The value of $t_{\rm c}$ is given in Table  \ref{tab:tempo_critico_all_modes}.

Therefore, the OMASS4 was implemented using the aforementioned packages, being applied for three different optimization modes: optimize SNR (mode~1), optimize AR (mode 2), and optimize both SNR and AR (mode 3). 

\begin{description}
    \item[Mode 1:] in this mode, the SNR is optimized, keeping the AR fixed. First, it is selected those modes that meet the AR requirement. Then, it is calculated the SNR value for each selected mode, using the maximum values for the $t_{\rm exp}$ and $G_{\rm em}$. The optimum mode is given by that one with the highest SNR.
    
    \item[Mode 2:] in this mode, the AR is optimized, keeping the SNR fixed. Initially, for each mode, it is calculated the minimum $t_{\rm exp}$ value that meet the SNR requirement, for the maximum $G_{\rm em}$ allowed. For this calculation, it is considered the values $s = S/t_{\rm exp}$, in photons/s of the star; the $s_{\rm sky} = S_{\rm sky}/t_{\rm exp}$, in photons/pixel/s of the sky; and the $s_{\rm dc} = S_{\rm dc}/t_{\rm exp}$, in e-/pixel/s of the DC. So, the Eq. \ref{eq:SNR_2} can be rewritten as follows
    
    \begin{equation}
             \mathcal{S} = \frac{s \times t_{\rm exp}}{\{ s \; t_{\rm exp} \; N_{\rm F}^2  + \\
            n_{\rm p} [\ (s_{\rm sky} +  s_{\rm dc}) \; t_{\rm exp} \; N_{\rm F}^2 +  \\
            (\sigma_{\rm ADU} \; G/G_{\rm em})^2 ]\ \}^{1/2}}.
        \label{eq:SNR_3_a}    
    \end{equation}

    Rearranging the terms of the Eq. \ref{eq:SNR_3_a} and isolating $t_{\rm exp}$,
    
    \begin{equation}
             s^2 \; t_{\rm exp}^2 -\\
             \mathcal{S}^2 \; N_{\rm F}^2 \; [\ s + n_{\rm p} (s_{\rm sky} + s_{\rm dc}) ]\ \; t_{\rm exp}  -\\
             \mathcal{S}^2 \; n_{\rm p} \; (\sigma_{\rm ADU} \; G/G_{\rm em})^2= 0 
    \label{eq:texp_min}        
    \end{equation}

     The minimum $t_{\rm exp}$ of Eq. \ref{eq:texp_min} is given by its smallest non-negative root. Therefore, the optimum mode is given through the calculation of the AR of the selected modes for the minimum $t_{\rm exp}$.
    
    \item[Mode 3:] in this mode, both SNR and AR are optimized. Initially, it is selected those modes which meet the SNR and AR at the same time. The resulting list of modes is used to create the space of states of the Bayes optimization method. Then, it is calculated the maximum values $\mathcal{S}^{\rm M}$ and $\mathcal{A}^{\rm M}$ and the minimum values $\mathcal{S}^{\rm m}$ and $\mathcal{A}^{\rm m}$ of the SNR and AR  respectively. They are used in normalization of both parameters into the range between 0 and 1. So, the function to be optimized is given by the multiplication of the normalized SNR and AR $\mathcal{A}$ values for each operation mode:
    
    \begin{equation}
         f = \frac{\mathcal{S} - \mathcal{S}^{\rm m}}{\mathcal{S}^{\rm M} - \mathcal{S}^{\rm m}} \times \frac{\mathcal{A} - \mathcal{A}^{\rm m}}{\mathcal{A}^{\rm M} - \mathcal{A}^{\rm m}}.
    \label{eq:funcao_custo_SNR_FA}        
    \end{equation}
    
    Therefore, the optimum mode for the CCD will be given by the set of parameters obtained through the Bayes optimization method that maximizes the function given by Eq. \ref{eq:funcao_custo_SNR_FA}.
\end{description}

\subsection{Artificial images simulator} \label{subsec:simulator}

An artificial image simulator was developed to perform the evaluation of the OMASS4. This simulator reproduces a star image that would be acquired by a CCD camera. It uses a Gaussian 2D, implemented in Python Language by the Astropy library \cite{Astropy2018}, as the star point spread function:

\begin{equation}
    f_{\rm p}(\mathcal{X},\mathcal{Y})  = \\
    \mathcal{C} e^{- a(\mathcal{X} - \mathcal{X}_{\rm 0})^2 \; - \; b(\mathcal{X} - \mathcal{X}_{\rm 0}) \times (\mathcal{Y} - \mathcal{Y}_{\rm 0}) \; - \; c(\mathcal{Y} - \mathcal{Y}_{\rm 0})^2},\\
\label{eq:moffat}    
\end{equation}

\noindent where

\begin{equation}
    a = \frac{cos(\Theta)^2}{2\delta^2_{\rm x}} + \frac{sin(\Theta)^2}{2\delta^2_{\rm y}},
    \label{eq:a_gaussian2D}    
\end{equation}

\begin{equation}
    b = \frac{sin(2 \Theta)^2}{2\delta^2_{\rm x}} - \frac{sin(2 \Theta)^2}{2\delta^2_{\rm y}},
\label{eq:b_gaussian2D}    
\end{equation}

\begin{equation}
    c = \frac{sin(\Theta)^2}{2\delta^2_{\rm x}} + \frac{cos(\Theta)^2}{2\delta^2_{\rm y}}.
\label{eq:c_gaussian2D}    
\end{equation}

\noindent $f_{\rm p}(\mathcal{X},\mathcal{Y})$ is the star intensity in ADU, $\mathcal{C}$ represents the maximum amplitude in ADU, $\mathcal{X}$ and $\mathcal{Y}$ are the coordinates over the image in pixels, $\mathcal{X}_{\rm 0}$ and $\mathcal{Y}_{\rm 0}$ are the star coordinates in pixels, $\delta_{\rm x}$ and $\delta_{\rm y}$ are the standard deviation of the Gaussian in the directions $\mathcal{X}$ and $\mathcal{Y}$, respectively; $\Theta$ is the rotation angle of the Gaussian.

The images created by the simulator have 200~x~200 pixels, the center coordinates of the star was fixed in $(\mathcal{X}_{\rm 0}, \mathcal{Y}_{\rm 0}) = (100,100)$ pixels; it uses values $\delta_{\rm x} = \delta_{\rm y}$ = 3/$B_{\rm in}$ pixels, and $\Theta$ = 0. The maximum amplitude $\mathcal{C}$ for each mode is calculated using

\begin{equation}
    \mathcal{C} = \frac{\beta \times t_{\rm exp} \times G_{\rm em} \times B_{\rm in}^2}{G},
\label{eq:amplitude_estrela}    
\end{equation}

\noindent where $\beta$ simulates a constant photon flux over the CCD, and $B_{\rm in}$ is the pixels binning parameter. It uses $\beta$ = 2000 photons/s. Over this image, another image was added with a background level $L_{\rm b}$, in ADU, and a Gaussian noise $N_{\rm b}$, in ADU/pixel. The $L_{\rm b}$ is calculated by 

\begin{equation}
    L_{\rm b} = B + (S_{\rm dc} + S_{\rm sky}) \times G_{\rm em} \times B_{\rm in}^2 / G.
\label{eq:nivel_bias}    
\end{equation}

\noindent $N_{\rm b}$ is calculated by 

\begin{equation}
     N_{\rm b} = \\
     \sqrt{(S_{\rm sky} +  S_{\rm dc}) \times \; (N_{\rm F} \; G_{\rm em} \; B_{\rm in})^2/ G  +  \sigma_{\rm ADU}^2 }.
\label{eq:ruido_imagem_artificial}    
\end{equation}

\noindent The values $S_{\rm dc}$, $G_{\rm em}$, $\sigma_{\rm ADU}$, $B_{\rm in}$, and $G$ are obtained as a function of the operation mode of the CCD; the $B$ value is obtained through a bias image; an arbitrary value for the $s_{\rm sky}$ of 12.3~photons/pixel/s was used.


\section{Results} \label{sec:results}

In order to facilitate the representation of the CCD operation modes, a codification for the parameters controlled by the OMASS4 was developed. This codification is presented in Table \ref{tab:codification}. Therefore, an operation mode can be represented by a sequence of 5 digits, organized in the same sequence as the parameters presented in Table \ref{tab:codification}. Thus, sequence 11111, for example, corresponds to the operation mode conventional, HSS~=~0.1 MHz, PA~=~1, Bin~=~1 pixel, and SI~=~256~x~256 pixels. However, to represent the operation modes of the images created through the simulator, a sequence with 4 digits was used. The last digit for the SI parameter was omitted, given that these images have a fixed size of 200 x 200 pixels.

\begin{table}[h!]
\centering
\caption{Codification used to represent the operation mode of the CCDs.}
\begin{tabular}{cccccccc}
\hline
Parameter & Unit  & \multicolumn{6}{c}{Code}  \\
\cline{3-8}
           &          & 1     & 2    & 3     & 4   & 5  &  6\\
\hline
EM Mode                 &          & Conv  & EM   &       &     &     &\\
HSS                     & (MHz)    & 0.1   & 1    & 3     & 10  & 20  & 30  \\
PA                      &          & 1     & 2    &       &     &     &\\
Bin                     & (pixels) & 1     & 2    &       &     &     &\\
SI                      & (pixels) & 256  & 512   &  1024 &     &     &\\
\hline
\end{tabular}
\label{tab:codification}
\end{table}

\subsection{Testing the OMASS4} \label{subsec:software_tests}

The tests presented in this section aim to evaluate the convergence of the OMASS4 to the optimum operation mode, the validity of the mode obtained, and the performance of the optimum mode concerning a mean performance of the objective function. These tests were executed for the optimization modes 1, 2, and 3. In addition, the accuracy of the SNR calculation was evaluated.

\subsubsection{Convergence} \label{subsubsec:convergence}
This test aims to evaluate if the OMASS4 converges to the same optimum operation mode for a series of images acquired with different operation mode for a constant light source. This series of images was generated using the simulator presented in Sec.~\ref{subsec:simulator}. For each image, the OMASS4 was run using the minimum values for the AR, and SNR of 2~fps and 100, respectively. 170 iterations were used in the Bayes optimization method execution and a maximum value for the star radius of 20 pixels. Table \ref{tab:convergence_optimum_mode} presents the values for the optimization of the SNR, AR, and SNR $\times$ AR, as well as, the modes used to generate each image. For each optimization mode, the operation mode of the CCD was the same for the entire series. Table \ref{tab:optimum_modes_convergence} presents the optimum operation modes obtained for each optimization mode, as well as, the mean value of the objective function for the results presented in Table \ref{tab:convergence_optimum_mode}. So, it was possible to demonstrate that the OMASS4 converges to the same optimum mode for an image series acquired with different operation modes, but for a constant light source.

\begin{table}[h!]
\centering
\caption{Optimum values obtained by the optimization of the SNR, AR and SNR $\times$ AR for a series of images generated with different operation modes.}
\begin{tabular}{cccccc}
\hline
 Operation Mode & $t_{\rm exp}$ & $G_{\rm em}$ & SNR    & AR    & SNR $\times$ AR \\
    &    (s)    &     &    &   (fps)    &        \\
\hline
1111 & 20 & 1  & 207.7 & 6.77 & 0.15 \\
1211 & 15 & 1  & 207.6 & 6.77 & 0.15 \\
1211 & 20 & 1  & 207.7 & 6.78 & 0.15 \\
1211 & 30 & 1  & 207.6 & 6.77 & 0.15 \\
1212 & 20 & 1  & 207.5 & 6.76 & 0.15 \\
1221 & 20 & 1  & 207.6 & 6.77 & 0.15 \\
2211 & 20 & 2  & 207.9 & 6.78 & 0.15 \\
2211 & 20 & 5  & 207.6 & 6.77 & 0.15 \\
2211 & 20 & 10 & 207.6 & 6.77 & 0.15 \\
2211 & 20 & 15 & 207.6 & 6.77 & 0.15 \\
2211 & 20 & 20 & 207.8 & 6.78 & 0.15 \\
2411 & 20 & 2  & 207.6 & 6.77 & 0.15 \\
2511 & 20 & 2  & 207.5 & 6.76 & 0.15 \\
2611 & 20 & 2  & 207.8 & 6.78 & 0.15 \\
\hline
\end{tabular}
\label{tab:convergence_optimum_mode}
\end{table}

\begin{table}[h!]
    \centering
    \caption{Optimum operation modes obtained through the execution of the OMASS4 for a series of images generated with different operation modes, for each optimization mode. The fifth column contains the mean values obtained for the objective function calculated using the values presented in Table \ref{tab:convergence_optimum_mode}}.
    \begin{tabular}{ccccc}
    \hline
    Optimization Mode & Operation Mode & $t_{\rm exp}$ & $G_{\rm em}$ & Objective \\
    & & (s)  &  & Function \\
    \hline
    1  & 12222 & 0.5  & 1   & 207.64 $\pm$  0.07\\
    2  & 12221 & 0.1477 $\pm$ 1e-4 & 1 & 6.771 $\pm$ 4e-3\\
    3  & 12221 & 0.1481 $\pm$ 1e-4 & 1 & 0.15241 $\pm$ 4e-5\\
    \hline
    \end{tabular}
    \label{tab:optimum_modes_convergence}
\end{table}

\subsubsection{Validation of the optimum mode} \label{subsubsec:opt_mode_validation}

\begin{figure}[h!]
    \centering
    \includegraphics[scale = 0.65]{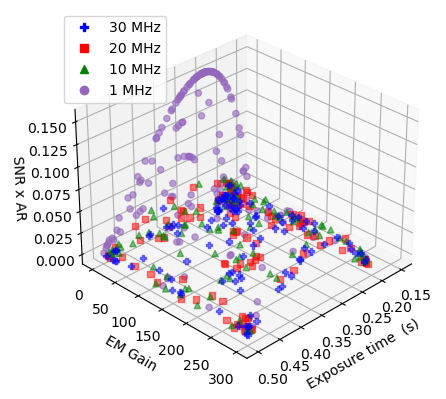}
    \caption{Optimization of the SNR $\times$ AR as a function of the $t_{\rm exp}$ and the $G_{\rm em}$, over the Bayes optimization method iterations. Modes HSS 1 MHz (purple), 10 MHz (green), 20 MHz (red), and 30 MHz (blue) are shown}.
    \label{fig:BOM_iterations}
\end{figure}

This test aims to evaluate if the operation mode returned by the OMASS4 corresponds to the global maximum. For this reason, an image was generated using the simulator with the operation mode 2211, $t_{\rm exp}$ = 1 s, and $G_{\rm em}$~=~20. The OMASS4 was run on this image. The values for the AR and the SNR of 2~fps and 100 were respectively used. Again, 170 iterations were used in the BOM execution and a maximum value for the star radius of 20 pixels. Table \ref{tab:optimum_modes_validation} presents the optimum mode obtained for each optimization mode. For the validation of the optimization mode 1, an image was generated with the optimum mode. The SNR calculated for this image is 208. Then, a series of images was generated with all operating modes that meet the AR requirement, for the maximum values of $t_{\rm exp}$ and $G_{\rm em}$. For each image, the SNR value was calculated. For the validation of the optimization mode 2, the AR was calculated for all modes that meet the SNR requirement, for the minimum $t_{\rm exp}$ value. For the validation of the optimization mode 3, the objective function was evaluated for 500 iterations for all modes that meet both SNR and AR requirements, for random values of $t_{\rm exp}$ and $G_{\rm em}$. Tables \ref{tab:SNR_setups_around} to \ref{tab:SNR_AR_setpus_around} presents the 10 best values for the objective function for the optimization mode 1, 2, and 3, respectively. None of the presented modes achieved a better result than its respective optimum mode. For the optimization mode 2, 3 modes were obtained with the same AR. In this case, the OMASS4 selected the one with the larger SI. Figure \ref{fig:BOM_iterations} presents the SNR $\times$ AR values obtained as a function of the $t_{\rm exp}$, $G_{\rm em}$ and HSS over the Bayes optimization method iterations. Examining this figure, it is possible to see a maximum point for the HSS~=~1~MHz. However, it should be highlighted that the profile of the SNR $\times$ AR function varies as a function of the scientific requirements for SNR and AR. Therefore, there may be some cases where the optimum point would not be so defined as presented in Figure \ref{fig:BOM_iterations}.

\begin{table}[h!]
    \centering
    \caption{Optimum modes of the CCD obtained through the execution of the OMASS4 over an image generated with the operation mode 2211, for each optimization mode.}
    \begin{tabular}{@{}ccccc}
    \hline
    Optimization Mode  & Operation Mode & $t_{\rm exp}$ & $G_{\rm em}$ & Objective function\\
    &  &  (s) &  &  \\
    \hline
    1  & 12211 & 0.50  & 1  & 208 \\
    2  & 12221 & 0.15  & 1  & 6.78 \\
    3  & 12221 & 0.25  & 1 & 0.152 \\
    \hline
    \end{tabular}
    \label{tab:optimum_modes_validation}
\end{table}

\begin{table}[h!]
    \centering
    \caption{10 best values obtained for the SNR for all operation modes that meet the AR requirement. The $G_{\rm em}$ and $t_{\rm exp}$ values were fixed in 300x (for the EM modes) and 0.5 s, respectively.}
    \begin{tabular}{@{}cccc}
    \hline
    Operation Mode  & $t_{\rm exp}$ & $G_{\rm em}$ & SNR    \\
    & (s) & & \\
    \hline
    12123 & 0.5          & 1           & 194 \\
    12213 & 0.5          & 1           & 179 \\
    12113 & 0.5          & 1           & 156 \\
    26123 & 0.5          & 300         & 156 \\
    24123 & 0.5          & 300         & 156 \\
    22113 & 0.5          & 300         & 156 \\
    24213 & 0.5          & 300         & 156 \\
    26223 & 0.5          & 300         & 156 \\
    25223 & 0.5          & 300         & 156 \\
    22123 & 0.5          & 300         & 156 \\
    \hline
    \end{tabular}
    \label{tab:SNR_setups_around}
\end{table}

\begin{table}[h!]
    \centering
    \caption{10 best values obtained to the AR for all operation modes that meet the SNR requirement.}
    \begin{tabular}{@{}cccc}
    \hline
    Operation Mode  & $t_{\rm exp}$ & $G_{\rm em}$ & AR   \\
    & (s) & &\\
    \hline
    12121 & 0.18     & 1        & 5.56 \\
    24221 & 0.20     & 300      & 4.91 \\
    24223 & 0.20     & 300      & 4.91 \\
    24222 & 0.20     & 300      & 4.91 \\
    25222 & 0.20     & 300      & 4.91 \\
    25223 & 0.20     & 300      & 4.91 \\
    25221 & 0.20     & 300      & 4.91 \\
    22221 & 0.20     & 300      & 4.91 \\
    22121 & 0.20     & 300      & 4.91 \\
    24213 & 0.20     & 300      & 4.91 \\
    \hline
    \end{tabular}
    \label{tab:AR_setups_around}
\end{table}

\begin{table}[h!]
    \centering
    \caption{10 best values for the SNR $\times$ AR obtained for all modes that meet both SNR and AR requirement. Note results were obtained after 500 iterations. Random values were selected for $G_{\rm em}$ and $t_{\rm exp}$.}
    \begin{tabular}{ccccc}
    \hline
    Operation Mode & $t_{\rm exp}$ & $G_{\rm em}$ & SNR $\times$ AR   \\
    & (s) &  &\\
    \hline
    12221 & 0.23      & 1        & 0.145           \\
    12222 & 0.30      & 1        & 0.141           \\
    12221 & 0.30      & 1        & 0.141           \\
    12222 & 0.26      & 1        & 0.130           \\
    12221 & 0.34      & 1        & 0.120           \\
    12222 & 0.34      & 1        & 0.120           \\
    12121 & 0.29      & 1        & 0.109           \\
    12122 & 0.29      & 1        & 0.109           \\
    12221 & 0.19      & 1        & 0.106           \\
    12121 & 0.32      & 1        & 0.104           \\
    \hline
    \end{tabular}
    \label{tab:SNR_AR_setpus_around}
\end{table}

\subsubsection{Optimization Performance} \label{subsubsec:opt_performance}

This test aims to evaluate the performance obtained by the optimum mode in comparison with a mean performance of the objective function over the allowed operation modes of the CCD. Initially, a series of images was created with the operation mode 1211, varying the incident light flux from $\beta$~=~2000~photons/s to $\beta$~=~3000~photons/s with steps of $\beta$~=~100~photons/s. The OMASS4 was run on each of the images. The used values for the AR and the SNR was 0.05 fps and 100, respectively. 170 iterations were used for the Bayes optimization method and a maximum value for the star radius of 20 pixels. The mean value of the objective function was calculated for 500 random iterations over the operation modes that meet the requirements of SNR and/or AR. Table \ref{tab:performance} presents the results obtained for this experiment. $R_{\rm OMASS4}$ is the maximum value of the objective function obtained by running the OMASS4 over the images. $R_{\rm OF}$ represents the mean value of the objective function for 500 random iterations. Both $R_{\rm OMASS4}$ and $R_{\rm OF}$ are presented for the three optimization modes. Also, tabulated is the relative percentage difference $R$, where $R~=~(R_{\rm OMASS4}/R_{\rm OF} - 1) \times 100$. Using these results, it is possible to conclude that the performance obtained through the OMASS4 was better than the performance for a mean value of the objective function. This result presents the usefulness of the OMASS4 for an observer with no knowledge about the CCDs operation. The best results obtained for the optimization modes 1, 2, and 3 were R~=~265~\%, R~=~428~\%, and R~=~165~\%, respectively.

\begin{table*}[h!]
\begin{center}
\caption{Objective function values obtained for a series of images with different incident light flux, for each optimization mode. The results $R_{\rm OMASS4}$ obtained by running the OMASS4 over the images} and the mean value of the objective function $R_{\rm OF}$ for 500 iterations, over the allowed operation modes, are presented. Additionally shown is the the relative percentage difference $R$, in \%, where $R = (R_{\rm OMASS4}/R_{\rm OF} - 1) \times 100$.
\begin{tabular}{cccccccccc}
\hline
$\beta$ & \multicolumn{2}{c}{SNR} & R & \multicolumn{2}{c}{AR (fps)} & R & \multicolumn{2}{c}{SNR $\times$ AR} & R \\
\cline{2-3} \cline{5-6} \cline{8-9}
(f/s)   & $R_{\rm OMASS4}$      & $R_{\rm OF}$     & (\%) & $R_{\rm OMASS4}$      & $R_{\rm OF}$     & (\%) & $R_{\rm OMASS4}$      & $R_{\rm OF}$     & (\%)     \\
\hline
2000 & 1395 & 402 & 247 & 6.8 & 1.3 & 417 & 0.024 & 0.01  & 140 \\
2100 & 1434 & 416 & 245 & 6.8 & 1.3 & 429 & 0.025 & 0.01  & 150 \\
2200 & 1472 & 433 & 240 & 6.9 & 1.4 & 385 & 0.025 & 0.01  & 150 \\
2300 & 1509 & 435 & 247 & 7.1 & 1.4 & 396 & 0.025 & 0.01  & 150 \\
2400 & 1546 & 442 & 250 & 7.1 & 1.5 & 390 & 0.025 & 0.01  & 150 \\
2500 & 1582 & 464 & 241 & 7.1 & 1.4 & 402 & 0.026 & 0.01  & 160 \\
2600 & 1616 & 472 & 242 & 7.2 & 1.6 & 361 & 0.026 & 0.01  & 160 \\
2700 & 1651 & 470 & 251 & 7.2 & 1.7 & 322 & 0.027 & 0.01  & 170 \\
2800 & 1684 & 461 & 265 & 7.2 & 1.7 & 336 & 0.027 & 0.01 & 145 \\
2900 & 1717 & 489 & 251 & 7.4 & 1.7 & 336 & 0.027 & 0.01 & 145 \\
3000 & 1749 & 503 & 248 & 7.7 & 1.7 & 353 & 0.027 & 0.01  & 170 \\
\hline
\end{tabular}
\label{tab:performance}
\end{center}
\end{table*}

\subsubsection{SNR calculation} \label{subsubsec:SNR_calculation}

This test aims to evaluate the accuracy of the SNR calculated by the OMASS4 for an image of the object. Initially, the theoretical value was calculated for an image series created using the simulator. For each image, the radius and the number of the pixels of the star were calculated, according to the FWHM of the star, as presented in Sec. \ref{sec:SNR}. The signal S of the star was obtained by the sum of the pixels with coordinates (x,y) within the star radius, given by the Eq. \ref{eq:moffat}. The values of $S_{\rm dc}$, $\sigma_{\rm ADU}$, $G$, $G_{\rm em}$, and $B_{\rm in}$ were obtained using the operation mode of the CCD. The same values for the $s_{\rm sky}$ presented in Sec. \ref{subsec:simulator} were used. Table \ref{tab:SNR_comparation} presents the used operation mode of the image series, the calculated value of the SNR by the OMASS4, and the theoretical value of the SNR. Thus, it is possible to conclude that the model used to calculate the SNR can represent the SNR value that would be obtained by acquiring an image of a star for different configurations of the CCD. The greatest difference between the theoretical and the calculated values was 1.05~\%.

\begin{table}[h!]
    \centering
    \caption{Comparation between the SNR values calculated (Calc) by the OMASS4 and the theoretical (Theor) for a series of images with different operation modes. Presented are the operation mode of each image, the obtained SNR values and the difference between them.}
    \begin{tabular}{ccccccccc}
    \hline
    Operation Mode & $t_{\rm exp}$ & $G_{\rm em}$ &  \multicolumn{2}{c}{SNR}  & Difference\\
    \cline{4-5}
    & (s) & & Calc & Theor & (\%)\\   
    \hline
    \hline
    1211 & 20 & 1  & 1377.6 & 1378.0 & 0.03  \\
    1212 & 20 & 1  & 1389.7 & 1393.9 & 0.30  \\
    1221 & 20 & 1  & 1386.3 & 1385.3 & -0.08 \\
    1222 & 20 & 1  & 1392.8 & 1394.5 & 0.12  \\
    2211 & 1  & 20 & 220.2  & 219.8  & -0.18 \\
    2212 & 1  & 20 & 220.6  & 221.0  & 0.17  \\
    2221 & 1  & 20 & 221.6  & 220.7  & -0.38 \\
    2222 & 1  & 20 & 221.3  & 221.2  & -0.05 \\
    2222 & 1  & 25 & 221.1  & 221.2  & 0.07  \\
    2411 & 1  & 20 & 210.5  & 210.2  & -0.16 \\
    2511 & 1  & 20 & 187.1  & 188.0  & 0.50  \\
    2611 & 1  & 20 & 147.0  & 148.6  & 1.04 \\
    \hline
    \end{tabular}
    \label{tab:SNR_comparation}
\end{table}

\subsection{Optimization of the observation nights} \label{subsec:OPTMize_observations}

This section presents the use of the OMASS4 introduced in this paper to optimize the performance of the CCDs in public data from observations obtained at the Observatório Pico dos Dias. The data from nights between April 2017 and April 2018 were analyzed.  More recent observations could not be used due to the proprietary time of 2 years~\cite{LNA_banco_dados}. The selected nights were those with observations of photometric time series of point-like sources carried out on the 1.6 m Perkin-Elmer telescope, and with the use of an iXon EMCCD. Nights which observed extended objects like clusters, nebulae, and galaxies were excluded. Table \ref{tab:selected_nights} presents the information for all selected nights. For all data sets, the astronomical objects were matched from their information in SIMBAD \cite{SIMBAD}. The duration of the time series for each night was obtained using the time interval between the first image and the last image of the series. The sum of the duration of all time series is 49 hours and 41 minutes. It is known that the scientific subject in the night of 03-06-2017 was stellar occultations. As we do not know what star in the field of view was the target, we just selected the brightest one to be optimized. The values of the SNR presented in Table \ref{tab:selected_nights} represent the mean calculated for the 10 highest values found in the series. The AR for each night was calculated by the duration of the time series divided by the number of exposures.

For each night, the OMASS4 was run for each optimization mode. The SNR and AR values presented in Table \ref{tab:selected_nights} were used. The SI and Bin were set to the same values as the ones used by the observer. 170 iterations for the Bayes optimization method were used. Tables \ref{tab:opt_modes_SNR} to \ref{tab:opt_modes_SNR_AR} present the results obtained for the optimization of the SNR, AR and SNR $\times$ AR, Tables \ref{tab:opt_modes_SNR} and \ref{tab:opt_modes_AR} present the the relative percentage difference $R$ between the value of the objective function and the respective value obtained by the observer. Table \ref{tab:opt_modes_SNR_AR} presents both the SNR and AR obtained by the OMASS4 and their respective relative percentage differences, $R_{\rm SNR}$ and $R_{\rm AR}$, compared to the observer's values. The AR values presented in Tables \ref{tab:opt_modes_AR} and Table \ref{tab:opt_modes_SNR_AR} take into account the overhead time between image cubes of the acquisition system presented in Sec. \ref{sec:acquisition_system}. This overhead time is 980 ms for each cube with 70 images. This overhead was added to the observation time needed to acquire each image series with the optimum AR. The ARs correspond to the total time calculated divided by the number of images in the series. 

\begin{table*}[h!]
    \begin{center}
    \caption{Information related to the selected observation nights performed at the Observatório Picos dos Dias for the execution of the OMASS4. Tabulated are the date of the observation, the used operation mode, the type of observed object (TOO), the number of exposes in the series $N_{\rm e}$, the mean SNR, the AR and the observation time. The following abbreviation for the objects are used: EB = eclipsing binary, WD = white dwarf, ST = star, and EX = exoplanet. The type of objected was found using the astronomical data bank SIMBAD \cite{SIMBAD}.}
    \begin{threeparttable}
    \begin{tabular}{@{}ccccccccc}
    \hline
    Night\tnote{1}    & Operation Mode & $t_{\rm exp}$ & $G_{\rm em}$ & TOO & $N_{\rm e}$ & SNR & AR & Duration \\
    (d-m-y)  & & (s) & & & & & (fps) & (h) \\
    \hline
    \hline
    22-03-17 & 12213 & 3   & 1 & EB & 1234 & 52.6 $\pm$ 1.3    & 0.076  & 4.5 \\
    23-03-17 & 12213 & 2   & 1 & EB & 1692 & 34.6 $\pm$ 0.6    & 0.091  & 5.1 \\
    24-03-17 & 12213 & 2   & 1 & EB & 1319 & 35.6 $\pm$ 1.5    & 0.11   & 3.3 \\
    14-04-17 & 11223 & 40  & 1 & WD & 134  & 49.9 $\pm$ 1.7    & 0.014  & 2.7 \\
    15-04-17 & 11223 & 20  & 1 & WD & 414  & 67.8 $\pm$ 2.6    & 0.019  & 6.2 \\
    16-04-17 & 22123 & 41  & 1 & WD & 341  & 94.5 $\pm$ 3.4    & 0.019  & 5.0 \\
    03-06-17 & 22113 & 100 & 1 & ST & 20   & 3964.4 $\pm$ 53.6 & 0.01   & 0.6 \\
    02-07-17 & 22123 & 45  & 1 & ST & 221  & 110.6 $\pm$ 8.9   & 0.02   & 3.0 \\
    13-07-17 & 12113 & 30  & 1 & EX & 811  & 1257.6 $\pm$ 1.6  & 0.032  & 7.1 \\
    08-08-17 & 12113 & 10  & 1 & EX & 1150 & 1097.0 $\pm$ 2.5  & 0.053  & 6.0 \\
    24-11-17 & 13123 & 15  & 1 & ST & 209  & 88.8 $\pm$ 4.3    & 0.056  & 1.0 \\
    06-03-18 & 22123 & 180 & 1 & WD & 63   & 63.8 $\pm$ 4.1    & 0.0035 & 5.0 \\
    \hline
    \end{tabular}
    \begin{tablenotes}
    \item[1] The dates presented in this table are formatted in the same format used by the observatory, and they correspond to the start of the observation night.
    \end{tablenotes}
    \end{threeparttable}
    \label{tab:selected_nights}
    \end{center}
\end{table*}

\begin{table}[h!]
    \centering
    \caption{Operation modes of the CCDs obtained through the SNR optimization of the observation nights at the Observatório Picos dos Dias. Tabulated is the SNR obtained by the OMASS4 and its relative percentage difference $R$ with respect to the SNR of the operation mode used by the observer.}
    \begin{tabular}{ccccccc}
    \hline
    Night    & Operation Mode & $t_{\rm exp}$ & $G_{\rm em}$ & SNR & R \\
    (d-m-y) &    & (s)  &  &  & (\%)     \\
    \hline
    22-03-17 & 12213 & 13  & 1 & 131.6  & 145.3 \\
    23-03-17 & 11213 & 11  & 1 & 106.3  & 207.3 \\
    24-03-17 & 12213 & 9   & 1 & 96.4   & 170.9 \\
    14-04-17 & 11223 & 73  & 1 & 102.1  & 104.5 \\
    15-04-17 & 11223 & 54  & 1 & 168.3  & 148.4 \\
    16-04-17 & 11223 & 53  & 1 & 196.0  & 107.5 \\
    03-06-17 & 11213 & 104 & 1 & 7073.2 & 78.4  \\
    02-07-17 & 11223 & 49  & 1 & 165.8  & 49.9  \\
    13-07-17 & 11213 & 31  & 1 & 1297.0 & 3.1   \\
    08-08-17 & 11213 & 19  & 1 & 2189.7 & 99.6  \\
    24-11-17 & 11223 & 18  & 1 & 139.1  & 56.8  \\
    06-03-18 & 11223 & 283 & 1 & 114.6  & 79.5 \\
    \hline
    \end{tabular}
    \label{tab:opt_modes_SNR}
\end{table}

\begin{table}[h!]
    \centering
    \caption{Optimum operation modes of the CCDs obtained through the AR optimization of the observation nights at the Observatório Picos dos Dias. Tabulated is the AR value calculated by the OMASS4 and its relative percentage difference $R$ with respect to the AR obtained by the observer.}
    \begin{tabular}{ccccccc}
    \hline
    Nights    & Operation Mode & $t_{\rm exp}$ & $G_{\rm em}$  & AR & R \\
    (d-m-y) &    & (s) &  & (fps) & (\%)     \\
    \hline
    22-03-17 & 12213 & 3  & 1   & 0.3  & 324.1 \\
    23-03-17 & 12213 & 2  & 1   & 0.5  & 419.7 \\
    24-03-17 & 24213 & 2  & 300 & 0.5  & 361.7 \\
    14-04-17 & 11223 & 17 & 1   & 0.1  & 316.1 \\
    15-04-17 & 11223 & 9  & 1   & 0.1  & 513.1 \\
    16-04-17 & 11223 & 13 & 1   & 0.1  & 318.1 \\
    03-06-17 & 11213 & 33 & 1   & 0.03 & 217.8 \\
    02-07-17 & 11223 & 22 & 1   & 0.04 & 124.3 \\
    13-07-17 & 11213 & 29 & 1   & 0.03 & 6.3   \\
    08-08-17 & 11213 & 5  & 1   & 0.2  & 295.3 \\
    24-11-17 & 11223 & 8  & 1   & 0.1  & 133.8 \\
    06-03-18 & 11223 & 88 & 1   & 0.01 & 222.0 \\
    \hline
    \end{tabular}
    \label{tab:opt_modes_AR}
\end{table}

\begin{table*}[h!]
    \begin{center}
    \centering
    \caption{Operation modes of the CCDs obtained through the SNR $\times$ AR optimization of the observation nights at the Observatório Picos dos Dias. Tabulated are the SNR and AR values calculated by the OMASS4 and their relative percentage difference R$_{\rm SNR}$, and R$_{\rm AR}$ with respect to the values obtained by the observer.}
    \begin{tabular}{cccccccc}
    \hline
    Night    & Operation Mode & $t_{\rm exp}$ & $G_{\rm em}$ & SNR & R$_{\rm SNR}$ & AR & R$_{\rm AR}$ \\
    (d-m-y) &   & (s) &  &  & (\%) & (fps) & (\%)     \\
    \hline
    22-03-17 & 12213 & 6   & 1   & 79.6   & 48.3 & 0.17  & 124.3 \\
    23-03-17 & 12213 & 4   & 1   & 57.7   & 66.9 & 0.23  & 152.2 \\
    24-03-17 & 12213 & 2   & 267 & 57.4   & 61.3 & 0.25  & 123.7 \\
    14-04-17 & 11223 & 33  & 1   & 68.7   & 37.6 & 0.03 & 120.6 \\
    15-04-17 & 11223 & 19  & 1   & 100.1  & 47.8 & 52.0  & 181.9 \\
    16-04-17 & 11223 & 24  & 1   & 131.1  & 38.8 & 42.0  & 121.0 \\
    03-06-17 & 11213 & 55  & 1   & 5168.7 & 30.4 & 18.0  & 87.1  \\
    02-07-17 & 12223 & 31  & 1   & 133.8  & 20.9 & 31.0  & 53.5  \\
    13-07-17 & 11213 & 30  & 1   & 1277.0 & 1.5  & 33.0  & 3.1   \\
    08-08-17 & 11213 & 9   & 1   & 1498.4 & 36.6 & 0.11  & 112.7 \\
    24-11-17 & 11223 & 11  & 1   & 109.7  & 23.6 & 88.0  & 57.5  \\
    06-03-18 & 12123 & 150 & 1   & 83.47  & 30.8 & 0.007 & 88.4 \\
    \hline
    \end{tabular}
    \label{tab:opt_modes_SNR_AR}
    \end{center}  
\end{table*}

Based on these results, it is possible to conclude that we were able to obtain the optimum mode of the CCDs, according to the scientific requirements of SNR and AR, for each night, for each optimization mode. For the optimization of the SNR, it was possible to obtain an improvement of up to 207.3~\% in the SNR. In many cases, the exposure times given by the OMASS4 were longer than those set by the observer. This occurs because the overhead time between image cubes given by the acquisition system allows the use of a greater $t_{\rm exp}$, keeping the same AR of the observations. In this analysis, it was possible to observe different operation modes used by the observer, in contrast with the same operation mode returned by the OMASS4. This shows how the observers have different criteria in choosing the operation mode of the CCD. It was noticed that in some cases the EM mode is used without setting a $G_{\rm em}$ amplification greater than one. This results in a noisier performance of the CCD without the advantage of signal amplification given by the EM mode. Also, it was noticed some cases where it would be possible to use an HSS with a greater readout time (therefore, with a smaller read noise) without reducing the AR. For the AR optimization, an improvement in the AR was obtained of up to 513.1~\% for one of the cases. It was noticed that several optimum modes given by the OMASS4 results in a smaller HSS, despite having a higher AR. This occurs because the use of higher HSSs results in higher read noises. A greater $t_{\rm exp}$ is needed to achieve the same SNR given by the OMASS4. A suggestion in using the cameras is to consider the use of a Bin~=~2~pixels whenever possible. This option improves both the SNR and the AR, allowing the use of HSSs with larger readout times. For the optimization of the SNR and the  AR, it was possible to improve both parameters for each night. In this result, an improvement was obtained of up to 66.9~\% for the SNR and up to 181.9~\% for the AR,

\subsubsection{Telescope time saving}
The amount of telescope time that would be saved was analyzed using the optimum mode suggested by the OMASS4 for each night presented in Sec. \ref{subsec:OPTMize_observations}. It is important to note that this analysis assumes that observation times can be reduced indefinitely, which is probably not true in case of studies of stellar variability.  

\begin{figure}[h!]
    \centering
    \includegraphics[scale=0.55]{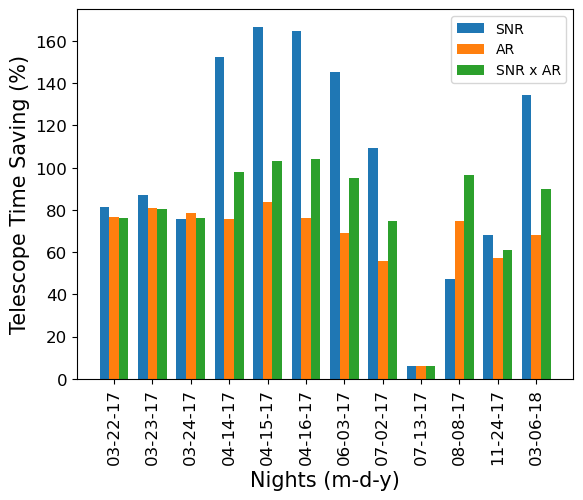}
    \caption{Telescope time saving values presented in Table \ref{tab:night_results_analysis}. Each triplet in the graph presents the value obtained for the optimization modes SNR (blue), AR (orange), and SNR x AR (green), respectively.}
    \label{fig:telescope_time_saving}
\end{figure}

This analysis was divided into the improvement obtained in the SNR and the AR. For the SNR case, it was calculated the $t_{\rm exp}$ needed for a CCD with the observer's operation mode achieves the same SNR obtained by the optimum mode. The $t_{\rm exp}$ value was calculated using the Eq. \ref{eq:texp_min}. The values adopted for the star and sky fluxes are those to match the SNR presented in Table \ref{tab:selected_nights}. Thus, the time saving was calculated by the difference between the $t_{\rm exp}$ of both cases, multiplied by the number of exposures in the series. The calculation of the time saving for the AR can be done more directly. The operation mode given by the OMASS4 is the one with the best AR and the same SNR. So, the time saving can be calculated from the difference between the $t_{\rm exp}$ of these cases, multiplied by the number of exposures. Table \ref{tab:night_results_analysis} presents the results obtained for this analysis. The $t_{\rm exp}$ in optimization mode 1 and 3 represents the time needed for the observer's operation mode to achieve the SNR of the optimum mode. The time saved in optimization mode 3 is given by the sum of the $t_{\rm exp}$ needed for the observer's mode to achieve the SNR of the optimum mode with the time saving related to the AR difference. At the bottom of the Table \ref{tab:night_results_analysis}, it is presented the percentage of the total time that would be saved if the OMASS4 was used as a tool to determine the optimum operation mode of the CCDs. Figure \ref{fig:telescope_time_saving} depicts the values, the telescope time saved, presented in Table \ref{tab:night_results_analysis}.

 \begin{table}[h!]
    \setlength{\tabcolsep}{4.5pt}
    \begin{center}
    \caption{Telescope time saving (TTS) analysis obtained from the optimum mode suggested by the OMASS4 in comparison with the observer's mode. Presented are the $t_{\rm exp}$ for the optimization modes 1 and 3 needed for one image acquired with the observer’s mode which meet the same SNR obtained by the optimum mode. The time saving of the optimization mode 3 represents the sum of the time saved with the improvement of both the AR and the SNR. At the bottom of the table are presented the percentage of the total time that would be saved if the OMASS4 was used as a tool to determine the optimum operation mode of the CCDs.}
    \begin{tabular}{cccccc}
    \hline
             & \multicolumn{2}{c}{Opt. Mode 1} & Opt. Mode 2 & \multicolumn{2}{c}{Opt. Mode 3} \\
    \cline{2-3} \cline{5-6}
    Night    & $t_{\rm exp}$     & TTS         & TTS     & $t_{\rm exp}$      & TTS        \\
    (d-m-y)  & (s)           & (\%)       & (\%)   & (s)            & (\%)      \\
    \hline
    22-03-17 & 13.8  & 81.3  & 76.4 & 5.8   & 76.2  \\
    23-03-17 & 11.5  & 86.9  & 80.8 & 4.2   & 80.3  \\
    24-03-17 & 8.9   & 75.8  & 78.3 & 3.9   & 76.3  \\
    14-04-17 & 150.5 & 152.6 & 75.8 & 71.4  & 97.9  \\
    15-04-17 & 110.2 & 166.8 & 83.8 & 40.8  & 103.1 \\
    16-04-17 & 128.2 & 164.8 & 76.1 & 67.1  & 104.2 \\
    03-06-17 & 251.4 & 145.4 & 69.0 & 150.5 & 95.1  \\
    02-07-17 & 98.6  & 109.4 & 55.7 & 64.6  & 74.9  \\
    13-07-17 & 31.9  & 6.0   & 6.2  & 30.9  & 6.0   \\
    08-08-17 & 18.9  & 47.3  & 74.7 & 18.2  & 96.3  \\
    24-11-17 & 27.3  & 68.2  & 57.2 & 19.4  & 61.1  \\
    06-03-18 & 564.6 & 134.6 & 68.2 & 300.4 & 89.9 \\
    \hline
    Total &        & 97.2  & 65.1 &        & 77.7  \\
    \hline
    \end{tabular}
    \label{tab:night_results_analysis}
    \end{center}
\end{table}

For the optimization mode 1, it would be needed 97.2~\% more time, i.e. 48.8 h, to obtain the same SNR for all nights keeping the same observer's operation mode. The use of this optimization mode would improve the data quality obtained by the observers for each night. For the optimization mode 2, the time saving would be 65.1~\% (32.3~h), if the mode with the best AR was used keeping the same SNR. The use of this optimization mode would allow the observer to acquire the same number of exposures, for smaller observation time. This fact would allow a higher number of projects to be allocated at the Observatório Picos dos Dias. For optimization mode 3, the time saving would be 77.7~\% (38.6~h) if both parameters were optimized at the same time. So, image acquisition would be done with better quality, in a smaller time. 

\subsubsection{Maximum Acquisition Rates provided by the SPARC4}

In this section, it is presented an estimate of the maximum ARs that the EMCCDs of the SPARC4 will provide considering stars with different magnitudes. This analyses helps understanding the possible scientific applications with the SPARC4, based on the limitations of acquisitions rates as a function of stellar magnitude. For each magnitude, the photon flux of the star was obtained through the Pogson Equation \cite{kepler2000}

\begin{equation}
    m - m_h = -2.5 log \frac{S}{S_{\rm h}}.
\label{eq:pogson}    
\end{equation}

\begin{figure}[h!]
    \centering
    \includegraphics[scale=0.5]{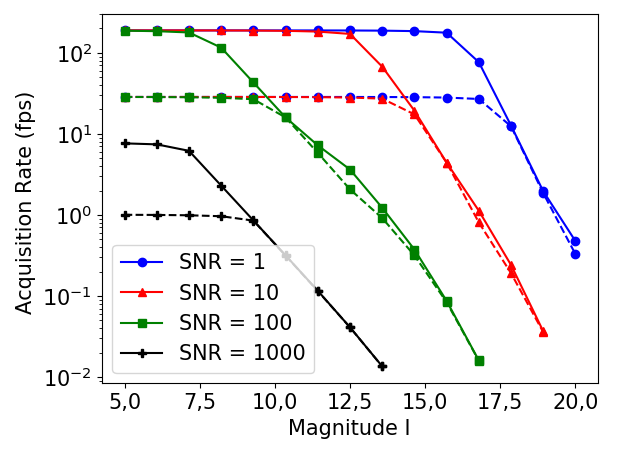}
    \caption{Acquisition rate of the CCD as a function of the star magnitude. Presented are the curves obtained for SNR 1 (blue), 10 (red), 100 (green), and 1000 (black). Solid lines represent the maximum values obtained for the acquisition rate  for the operation modes SI = (256, 512)~pixels and Bin = 2~pixels; dashed lines represented the maximum acquisition rate for the mode SI = 1024 pixels and Bin = 1 pixel.}
    \label{fig:AR_MAG}
\end{figure}

\noindent $m$ and $S$ represents the magnitude and the photon flux of the observed star. For this experiment, reference values of $m_{\rm h}$~=~12.25 \cite{SIMBAD} and $S_{\rm h}$~=~56122.30~photon/s were used. These are the magnitude and the star flux obtained for an arbitrarily selected image of the star HATS24 \cite{Oliveira_2019} in the I-filter, acquired using the 1.6 m telescope of the Observatório Pico dos Dias. In our analysis, magnitude values ranging from 5 to 20 were considered. Thus, the maximum ARs were obtained by executing the OMASS4 for each magnitude, for the SNR values of 1, 10, 100, and 1000. The OMASS4 execution was divided into two groups of operation modes: (1) SI~=~(256, 512) pixels and Bin~=~2~pixels; and (2) SI~=~1024~pixels and Bin~=~1~pixel. Figure \ref{fig:AR_MAG} presents the result obtained for this analysis. Solid lines represent the maximum values for the AR obtained for group 1; dashed lines represent the maximum values obtained for group 2. It can be seen a region with an AR of 189 fps, which is the highest AR allowed by the SPARC4 acquisition system. For the same value of SNR, there is a magnitude from which the AR of group 1 equals to the AR of group~2. At this point, it is indifferent to use modes with a smaller spatial resolution to achieve higher ARs. This happens when the $t_{\rm exp}$ required to maintain the SNR constant is larger than the readout time of a full-frame image. However, it should be highlighted that this result was done using the star magnitude in the filter I. Although, it is still advised to use an image of the object previously acquired (pre-image) to obtain the real performance of the CCD for the observed object in the observed band pass.

\section{Conclusion} \label{sec:conclusion}

We have presented the OMASS4 to optimize the parameters of an EMCCD to obtain the highest possible SNR and/or the fastest possible acquisition rate in astronomical observations. Our method is based on an empirical characterization of the CCD read noise and of the acquisition rate using a custom control system designed for the SPARC4 instrument. We applied the OMASS4 for the optimization of selected real observation nights, where we obtained an improved performance compared to that obtained with the operation mode selected by the observers. For the optimization modes~1 and 2, it was possible to obtain an improvement of up to 207.3~\%, and 513.1~\% for the SNR and AR, respectively. For the optimization mode 3, it was possible to obtain an improvement up to 66.9~\% for the SNR, and up to 181.9~\% for AR. We have also noticed that there would be a significant amount of telescope time saving if an optimization mode was used, with the assumption that observation times can be reduced indefinitely. For optimization mode 1, it was found that it would require 97.2~\% more time for the programs of the selected nights to achieve the same SNR obtained by the OMASS4. For optimization mode 2, it would be possible to save 65.1~\% of the total observation time, keeping the same SNR and number of exposures. For optimization mode 3, the optimization of both parameters would save 77.7~\% of the total observation time.

Among the selected nights it was not found a case that requires high sensitivity (SNR $>$ 100) and high acquisition rates (AR $>$ 10~Hz) at the same time. This would happen because of the lack of projects that use the Observatório Pico dos Dias's telescopes for this purpose. However, the SPARC4 instrument was designed to allow not only for the scientific subjects of the selected nights but also for other potential subjects involving the observation of objects that require high sensitivity and high ARs \cite{Rodrigues2012}. So, the use of the OMASS4 becomes even more important to help the observer in these situations. This fact contributes to the use of the observatory in carrying out research in subjects that was not possible until now.

In this project, the measurement of the characterization of the read noise and the AR was made only for one of the SPARC4 cameras. Although, these measurements would be repeated to the other SPARC4 cameras, as well as, for the iKon cameras of the observatory. Beyond the SI modes allowed by the SPARC4, a possible improvement would be to allow continuous SI values over the CCD chip. This change would allow optimizing the AR with a greater degree of freedom. A possible application for the presented method would be to integrate it with the acquisition system of the SPARC4. This implementation would allow the re-execution of the method during the night as a method to readjust the performance of the cameras, according to the climatic variations. Besides, the use of the OMASS4 is not restricted for the SPARC4 instrument. This method would be of great advantage for other CCD based instrumentation, like microscopy, with the appropriated adjust of the input parameters.

\subsection* {Acknowledgments} \label{sec:acknowledgment}

We want to acknowledge Janderson de Oliveira for providing us with the HATS-24\,b data. We also want to thank Fapesp (Proc. 2013/26258-4), CNPq (Proc. 303444/2018-5), Finep (Proc. 0/1/16/0076/00), FAPEMIG (Proc. APQ-00193-15), for their financial support, and CAPES (Proc. 88882.430116/2019-01) for the internship funding. E.M. acknowledges funding from the French National Research Agency (ANR) under contract number ANR-18-CE31-0019 (SPlaSH).


\bibliography{report}   
\bibliographystyle{spiejour}   


\section{Biographies} \label{sec:biographies}

\begin{itemize}
    \item Denis Varise Bernardes: graduated in Physics Engineering, Lorena School of Engineering (EEL - USP) from 2012 to 2017. Master student at the Federal University of Itajubá, from 2018 to 2020, with the project of the determination of the optimal modes of operation of the SPARC4 acquisition system. Ph.D. student in the Astrophysics course of the National Space Research Institute (INPE), with the project of the acquisition system of the SPARC4 instrument: acquisition software and image simulator.
    
    \item Eder Matioli: graduated in physics from the São Carlos Institute of Physics of the University of São Paulo in 2003. He completed his master's in 2006 and his Ph.D. in 2010 in astrophysics from the National Institute for Space Research, partially hosted at the University of Texas at Austin. He was a resident astronomer at the Canada-France-Hawaii Telescope from 2010 to 2013. Since 2013 he has been a Research Scientist at the Laboratório Nacional de Astrofísica and is currently a visiting postdoctoral researcher at the Institut d'Astrophysique de Paris. He has experience in optical and near-infrared astronomy, with an emphasis on exoplanets, brown dwarfs, and low-mass binaries.
    
    \item Danilo Henrique Spadoti: graduated in Electrical Engineering from the University of Itajubá (2002). Completed the Masters (2004) and Ph.D. (2008) in Electrical Engineering, with a major telecommunications at USP -University of São Paulo. In 2009, did postdoctoral studies abroad, in Nanophotonics Group’s, at Cornell University, Ithaca, USA. In 2010 receive a FAPESP post-doctoral scholarship at Mackenzie University. He is currently an Associate Professor in the area of telecommunications systems and applied electromagnetics at the Federal University of Itajubá-UNIFEI.
\end{itemize}

\end{document}